\documentclass[a4paper,fleqn,usenatbib]{mnras}

\usepackage{color}

\usepackage[T1]{fontenc}
\usepackage{ae,aecompl}
\usepackage{graphics}           

\usepackage{graphicx}	
\usepackage{amsmath, bm}	
\usepackage{amssymb}	
\usepackage{epstopdf}           

\title[The St{\o}rmer problem for an aligned rotator]{The St{\o}rmer problem for an aligned rotator}

\author[V.  Epp and O. N. Pervukhina]{
V. Epp,$^{1,2}$\thanks{E-mail: epp@tspu.edu.ru}
and O. N. Pervukhina$^{1}$
\\
$^{1}$Department of Theoretical Physics, Tomsk State Pedagogical University, Kievskaya 60,  Tomsk 634061, Russia\\
$^{2}$Department of Quantum Field Theory, Tomsk State University, Lenina 36, Tomsk 634050, Russia\\
}

\date{Accepted XXX. Received YYY; in original form ZZZ}

\pubyear{2017}

\begin{document}
\label{firstpage}
\pagerange{\pageref{firstpage}--\pageref{lastpage}}
\maketitle

\begin{abstract}
The effective potential energy of the particles in the field of rotating uniformly magnetized celestial body is investigated. The axis of rotation coincides with the axis of the magnetic field. Electromagnetic field of the body is composed of a dipole magnetic  and  quadrupole electric fields. The geometry of the trapping regions is studied as a function of the magnetic field magnitude and the rotation speed of the body. Examples of the potential energy topology for different values of these parameters are given. The main difference from the classical St{\o}rmer problem is that the single toroidal trapping region predicted by St{\o}rmer is divided into equatorial and off-equatorial trapping regions. Applicability of  the idealized model of a rotating uniformly magnetized sphere with a vacuum magnetosphere to real celestial bodies is discussed.
\end{abstract}




\section{Introduction}

The motion of charged particles in the field of a magnetic dipole was first studied by \citet{Stormer1907} and is described in detail in his book \citep{Stormer1955}. As is known, the equations of motion of a particle in a dipole magnetic field  can not be solved in terms of elementary functions. Detailed studies of the trajectories of charged particles in a dipole magnetic field, including methods of mathematical modelling, are described by Alfven in his book  \citep{Alfven1950}.
 Quasi-periodic trajectories in a dipole magnetic field, with generalization to an arbitrary axially symmetric magnetic field, are investigated by  \citet{Braun1970} and \citet{Shebalin2004}. A profound bibliography concerning the motion of charged particles in the Earth's magnetic field as of 1965 is given by   \citet{Dragt1965}. The motion of particles in the equatorial plane has been studied both numerically and analytically by many authors (e.g. \citet{Graef1938, Ioanoviciu2015, Dragt1965}). Analytical solutions of the equations of  motion  in the equatorial plane were obtained for the first time by  \citet{Graef1938}.

A rotating magnetized body induces an electric field inside and outside the body. Examples of such bodies are rapidly rotating planets with a relatively strong magnetic field and stars. In particular, extremely large electric fields are generated in the vicinity of neutron stars.
Early work in this field is devoted to the study of the motion of charged particles in the field of a conducting celestial body whose magnetic axis coincides with the axis of rotation, the so-called coaxial rotator. The first model of the magnetosphere of the coaxial rotator was proposed by  \cite{Goldreich1969ApJ}.
The Goldreich and Julian model was developed in numerous subsequent papers (see, for example,  review in  \citet{Michel1991}).
Among the first such articles we should note \cite{Michel1973, Jackson1976ApJ, Michel1980, Asseo1984MNRAS, Fitzpatrick}. The later studies are reviewed  in the recent papers by \citet{Karastergiou2014,  Chen2014}. 

	Another area of research is the magnetosphere of a magnetized body whose axis of rotation does not coincide with the magnetic axis (inclined rotator) \citep{Deutsch1955, Ostriker1969}. An extensive bibliography of works on this topic is given in \cite{Karastergiou2014}.
Much of this work is devoted to the study of the dynamics of a plasma rotating with a star and the calculation of the spatial distribution of charge in the magnetosphere.
The dynamics of charged particles was mostly simulated numerically  \citep{Chen2014, Wada2011, Sarychev2010}, or analysed qualitatively . The geometry of the effective potential energy of charged particles in the vacuum magnetosphere of an inclined rotator was investigated by \citet{Epp2013effective, EppMasterova2014}.
Extensive studies are devoted to the motion of charged dust particles in the field of a coaxial rotator taking into account the gravitational field and the electric field of a jointly rotating plasma  (see, for example \citet{Dullin2002, Inarrea2005,Jontof2012} and references therein). 

Thus, to date, there are detailed studies of the dynamics of charged particles either in a dipole magnetic field (Earth-like mag\-ni\-to\-spe\-re), or in the field of a coaxial rotator and a co-rotating plasma (pulsar-like mag\-ni\-to\-spe\-re). The motion of charged particles in a vacuum field of a coaxial rotator has not been studied sufficiently. The aim of this paper is to fill this gap. The problem is all the more relevant due to recent discovery of the aurora in the brown dwarf LSR J1835 + 3259 \citep{Hallinan2015}. 
As far as we know, only two papers \citep{Schuster1987, Thielheim1990} are devoted
 to the study of  trapping  regions of an aligned rotator. The allowed and forbidden regions  for particles with energy ${\cal E}> mc^2$ have been defined ($m$ is the rest mass and $c$ is the speed of light). 
 We investigate the effective potential energy of particles in the field of a coaxial rotator in more general case, in particular, for any values of ${\cal E}$.   
 
 Though  the electric field is involved, one should bear in mind that we assume a vacuum magnetosphere, i.e.  the induced quadrupole electric field is not as large as to pull the charged particles off the surface of the planet or star. Therefore, the results obtained are not applicable to a star magnetosphere  filled with plasma.

\section{Effective potential energy}
 The external magnetic field of a uniformly  magnetized sphere  with  a dipole moment  $\bm{\mu}$ is defined by the vector potential
\begin{equation}\label{eq1}
\bm A=\frac{\bm{\mu}\times\bm r}{r^{3}}.
\end{equation}

Rotation of the magnetized sphere in its own magnetic field induces either polarization of the  body material, or redistribution of the free charges in a conducting  body. Thus produced distribution of bound or free charges   creates a quadrupole electric field outside the body, with a potential determined by solution of the Laplace equation \citep{Landau_8}  
 \begin{equation}\label{w}
\varphi=\frac{\mu a^{2}\omega q}{3cr^{3}}\left(1-3\cos^{2}\theta \right),
\end{equation}
where $a$ is the  radius of the sphere, $\omega$ is the angular velocity of rotation, $q=(2\varepsilon+1)/(2\varepsilon+3)$, $\varepsilon$  is the body material permittivity. In the case of a conducting body $q=1$. Axis $z$ is directed along the magnetic moment $\bm \mu$, hence, $\mu$ is definitely positive. The direction of rotation is determined by the sign of $\omega$. Since most of the celestial bodies are conductive, we  assume that $q=1$. In case of dielectric sphere, one has to replace $\omega$ by $q\omega$  in all following equations.The conductivity of the body  material  does not matter, since we assume that there are no closed currents. 

We start with the relativistic Lagrangian \citep{Goldstein2002} of a particle of rest mass $m$ and electric charge $e$
\begin{equation}
L=\frac 12 mu_\nu u^\nu+\frac ec u_\nu A^\nu, 
\end{equation}
where $u^\nu=\mathring x^\nu$ is the four-dimensional velocity vector, and $x^\nu=(ct,r,\theta,\psi)$ is the four-dimensional radius-vector in the spherical coordinate system. The circle over the variable denotes  derivative with respect to the proper time,   $A^\nu$ is the four-dimensional potential of  electromagnetic field 
\begin{equation}
A^\nu=\frac{\mu}{r^3}\left[\frac{a^{2}\omega }{3c}\left(1-3\cos^{2}\theta \right), 0,0,1 \right].
\end{equation}
Since the  Lagrangian\footnote{
 $\mathring t$ is the derivative of the time with respect to the proper time, do not confuse with $i$.} 
\begin{align}
L=&\frac 12 m\left(c^2\mathring t^2 -\mathring r^2 -r^2\mathring\theta^2 - r^2\mathring\psi^2\sin^2\theta\right)\nonumber\\
&+\frac{e\mu a^2\omega}{3cr^3}\mathring t\left(1-3\cos^{2}\theta \right)-\frac{e\mu}{cr}\mathring\psi\sin^2\theta
\end{align}
 does not depend explicitly on time  and  coordinate $\psi$, the generalized momenta of the particle $p_0=\partial L/\partial \mathring x^0$ and $p_3=\partial L/\partial \mathring x^3$ are conserved. We denote the energy of the particle by ${\cal E}=cp_0$, and the angular momentum conjugate to the coordinate $x^3$, by $M=-p_3$ 
\begin{align}\label{p0}
{\cal E}&= mc^2\mathring t+\frac{e\mu a^2\omega}{3cr^3}(1-3\cos^2\theta),  \\
M&=\left(mr^2\mathring\psi +\frac{e\mu}{cr}\right)\sin^2\theta.
\label{p3}
\end{align}
We express $\mathring t$  from (\ref{p0}),  and $\mathring\psi$ from (\ref{p3})  and substitute these into the identity $u_\nu u^\nu=c^2$.  As a result we obtain
\begin{align}\label{e1}
m^2c^2(\mathring r^2+r^2\mathring\theta^2)&=\left[{\cal E}-\frac{e\mu a^2\omega}{3cr^3}(1-3\cos^2\theta) \right]^2\nonumber\\
&-\frac{c^2}{r^2\sin^2\theta}\left(M-\frac{e\mu}{cr}\sin^2\theta\right)^2 -m^2c^4.
 \end{align}
Thus, we have  reduced the problem to the motion in a two-dimensional space $(r,\theta)$.

 In the particular case $\omega=0$  the particle moves in a dipole magnetic field and its relativistic factor $\gamma=(1-v^2/c^2)^{1/2}$ is a conserved quantity. Hence, equation  (\ref{e1}) can be written in the form
 \begin{equation}
 \frac{{\cal E}^2-m^2c^4}{2m c^2}=\frac{m}{2}(\mathring r^2+r^2\mathring\theta^2)+\frac{1}{2m  r^2\sin^2\theta}\left(M-\frac{e\mu}{cr}\sin^2\theta\right)^2,
 \end{equation}
 where the left-hand side is an integral of motion. This equation corresponds to the motion of a particle in the St{\o}rmer potential    \citep{Stormer1955}
 \begin{equation}
 U(r,\theta)=\frac{1}{2m\gamma  r^2\sin^2\theta}\left(M-\frac{e\mu}{cr}\sin^2\theta\right)^2.
 \end{equation}
 
In order to define the relativistic potential,  we express $\cal E$ from equation (\ref{e1}) 
\begin{multline}
{\cal E}-\frac{e\mu a^2\omega}{3cr^3}(1-3\cos^2\theta)=\\
\sqrt{m^2c^4+m^2c^2(\mathring r^2+r^2\mathring\theta^2)+\frac{c^2}{r^2\sin^2\theta}\left(M-\frac{e\mu}{cr}\sin^2\theta\right)^2}.
\end{multline}
We decide for the plus sign in front of the square root, because this expression can be obtained directly from the general identity
\begin{equation}\label{epuls}
{\cal E}-e\varphi=\sqrt{m^2c^4+c^2\left( \bm P-\frac ec \bm A\right)^2},
\end{equation}
Since $\mathring r^2+r^2\mathring\theta^2\ge 0$, the variables $r$ and $\theta$ must obey the inequality 
\begin{multline}\label{energy}
{\cal E}\ge \sqrt{m^2c^4+\frac{c^2}{r^2\sin^2\theta}\left(M-\frac{e\mu}{cr}\sin^2\theta\right)^2}\\
+\frac{e\mu a^2\omega}{3cr^3}(1-3\cos^2\theta),
\end{multline}
which determines the regions allowed for the motion of a charged particle at given initial conditions in a two-dimensional space $(r,\theta)$.  It can be written in the form ${\cal E}\ge U$, where
\begin{multline}\label{U}
U= \sqrt{m^2c^4+\frac{c^2}{r^2\sin^2\theta}\left(M-\frac{e\mu}{cr}\sin^2\theta\right)^2}\\
+\frac{e\mu a^2\omega}{3cr^3}(1-3\cos^2\theta)
\end{multline}
plays the role of effective potential energy. 

We introduce the dimensionless coordinate $\rho=|\Gamma|r$, where the integral of motion $\Gamma$  has the dimension  of a reciprocal length
\begin{align}\label{l} 
\Gamma=\frac{Mc}{e\mu}.
\end{align}
 For the moment we assume that $M\neq 0$. The case $M=0$ will be considered in section~\ref{m_zero}. Next we introduce a dimensionless potential energy  $V=(U-mc^2)/mc^2$:
\begin{equation}\label{V}
V=\sqrt{1+P\left(\frac{1}{\rho\sin\theta}-\frac{\epsilon\sin\theta}{\rho^2}\right)^{2}}-1+\frac{\epsilon\Phi P}{3\rho^3}(1-3\cos^2\theta),
\end{equation}
with $\epsilon=\pm1$ being  the sign of $\Gamma$ which depends on  combination of signs of $e$ and  $M$, and
\begin{equation}
P=\frac{M^4}{m^2e^2\mu^2},\quad \Phi=\frac{a^2\omega m}{M}.
\end{equation}
We have extracted the rest energy $mc^2$ in order to assure the asymptotic $V\to 0$ as $\rho\to\infty$.   The inequality (\ref{energy}) with such defined potential takes the form 
\begin{equation}\label{ener1}
V\le \frac{{\cal E}-mc^2}{mc^2}.
\end{equation}
The constants $P$ and $\Phi$ are dimensionless. According to equation (\ref{V}), $\Phi$ specifies  the magnitude of   quadrupole potential,  and $P$ is responsible for the  particle angular momentum. One can see that $P$ is always positive, and $\Phi$ is the ratio of the angular momentum of a particle resting at the equator  to its initial angular momentum $M$.  The sign of factor $\epsilon\Phi$ coincides with the sign of the product $e\omega$. Hence, the potential energy of a particle is invariant  against the transformation $e\to-e,\,\omega\to-\omega,\,M\to-M$.

{
Let us examine the non-relativistic limit. As one can see from equation (\ref{epuls}), the second term under the radical sign for a non-relativistic particle is much less than the first one. Hence, we have to put $P\ll 1$ in equation (\ref{V}). The potential of the electric field in the vicinity of the celestial body must be also small enough, i.e. $e\varphi\ll mc^2$, or $e\mu\omega/mc^3a\ll 1$. On the other hand, the term responsible for the electric field in the effective potential  (\ref{V}) is of order $\Phi P/3R^3=e\mu\omega/3mc^3a$  on the surface of the body. Hence, the   non-relativistic limit is ensured by inequality $P\ll 1$. Expanding the potential (\ref{V}) into a power series in $P$ we obtain
\begin{equation}
V=\frac 12 P\left(\frac{1}{\rho\sin\theta}-\frac{\epsilon\sin\theta}{\rho^2}\right)^{2}+\frac{\epsilon\Phi P}{3\rho^3}(1-3\cos^2\theta),
\end{equation}
The non-relativistic effective potential may be written as
\begin{equation}\label{Vn}
V_n=\frac{V}{P}=\frac 12 \left(\frac{1}{\rho\sin\theta}-\frac{\epsilon\sin\theta}{\rho^2}\right)^{2}+\frac{\epsilon\Phi }{3\rho^3}(1-3\cos^2\theta).\end{equation}
The same expression can be derived from the non-relativistic Lagrangian
\begin{equation}
L=\frac{mv^{2}}{2}-e\phi+\frac{e}{c}\left(\bm{v}\bm{A}\right) ,
\end{equation}
where $\bm v$ is the particle velocity. The first term in potential (\ref{Vn}) is usually used for solution of the St{\o}rmer problem for a dipole magnetic field. The second term represents the contribution of  the electric field to potential energy. In the case $\omega = 0 \,\,  (\Phi = 0)$ equation  (\ref{Vn}) is an exact expression for the potential energy in the sense that it is valid for relativistic particles, but the mass $m$ must be replaced by $\gamma m$. In this case potentials  (\ref{Vn}) and  (\ref{V}) are equivalent -- they produce the same equations of motion.

%
\section{Qualitative analysis}\label{qual}
%
Let's consider how  rotation of a magnetized body affects its potential energy. In the case $\omega = 0 \,\, (\Phi = 0) $ we have the classical St{\o}rmer problem. The effective potential $V (r, \theta)$ has a rather simple shape. If the initial angular momentum  $ M$ is negative $(\epsilon=-1)$, then it is of a repulsive nature and the particle performs infinite motion with any initial parameters. If $ M $ is positive, then there is a single potential valley whose bottom coincides with the magnetic field line $\rho = \sin ^ 2 \theta $, and the potential $ V $ at the bottom of the valley is everywhere zero. In this sense, the bottom of the valley does not have a slope. This valley has a pass to the open region at the point $\rho = 2, \, \theta = \pi / 2 $. The dependence of $ V (R) $ in the equatorial plane  in cylindrical coordinates $R=\rho\sin\theta$ and $z=\rho\cos\theta$ and a typical trajectory are shown in Fig. \ref {Vster}.
\begin{figure}
\centerline{\includegraphics[width=2.8in]{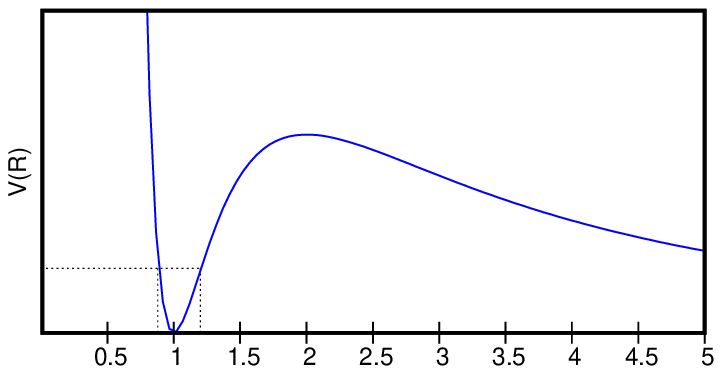}}
\centerline{\includegraphics[width=2.8in]{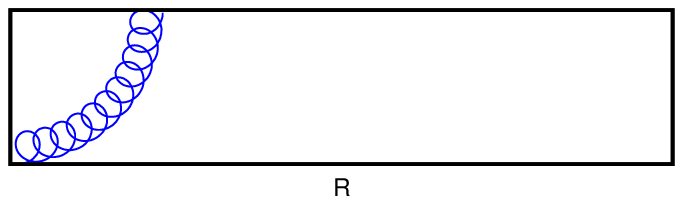}}
\caption{The St{\o}rmer's potential and a particle trajectory  in the equatorial plane.}
\label{Vster}
\end{figure}
The loop-like trajectory is the consequence of the angular momentum conservation (\ref{p3}).  In the dimensionless coordinates it can be written as ($M\neq 0)$
\begin{equation}\label{dotps}
\mathring{\psi}=\frac{M^3c^2}{me^2\mu^2\rho^2}\left(\frac{1}{\sin^2\theta}-\frac{1}{\rho}\right).
\end{equation}
One can see that the direction of the azimuthal motion of the particle changes each time as  the particle crosses the bottom of the potential valley $\rho = \sin ^ 2 \theta $.

If the magnetized  body is set to rotation, then the induced electric field shifts the bottom of the potential valley with respect to the line $\rho = \sin ^ 2 \theta $ in the direction $e\bm E$ \footnote{The sign of the radial  component of  electric field coincides with the sign of the  product $e \omega $ which, in turn, coincides with the sign of  $\epsilon\Phi$.}. Besides, the bottom line of the potential valley acquires a slope due to the dependence  on $\theta$.  These effects are described in detail in \cite{Schmidt1979}.

If, for example,  the   angular velocity vector coincides with the vector $\bm\mu$,  then the electric field in the equatorial plane points from the axis of rotation. It repels the positively charged particles and attracts negative ones. In this case, the potential of the electric field (\ref{w}) raises the bottom of the potential vally of Fig. \ref{Vster} and shifts it to the right as shown in Fig. \ref{four1}(a).
\begin{figure}
  \centerline{\includegraphics[width=\columnwidth]{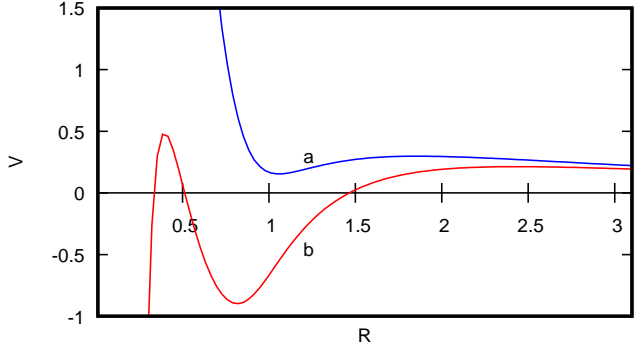}}
  \caption{Typical plots  of the potential $V (R)$ in the equatorial plane for $\epsilon=1$: $\Phi>0$ (a) and $\Phi<0$
  (b). {
  The curve (a) has minimum only if $\Phi<3-2\sqrt{2}$}.
  }
\label{four1}
\end{figure}
In the opposite case $\epsilon\Phi<0$ the electric field shifts the bottom of the potential valley toward the axis as shown in Fig. \ref{four1}(b) and Fig. \ref{four2}(a).
\begin{figure}
  \centerline{\includegraphics[width=\columnwidth]{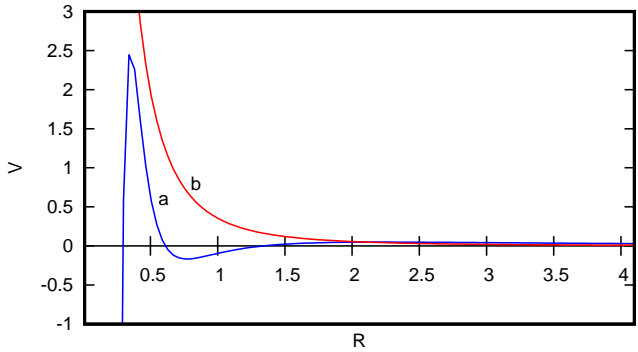}}
  \caption{Typical plots  of the potential $V (R)$ in the equatorial plane for $\epsilon=-1$: $\Phi>0$ (a) and $\Phi<0$
  (b). In each case, the plots with the largest possible number of stationary points are shown. {
  The curve (a) has minimum only if $\Phi>4\sqrt{2}$}.}
\label{four2}
\end{figure}
Besides, an infinitely deep potential well emerges in the centre.
A sufficiently strong positive electric potential raises the minimum of the St{\o}rmer potential so that the minimum of potential energy disappears (curve (b) in Fig.~\ref{four2}).  

The particle trajectories are changing respectively. Consider motion in the equatorial plane within a potential well. Particles of sufficiently low energy move around the bottom of the potential well and do not reach the circle of radius $\rho=1$. Hence, their azimuthal velocity will not change the sign -- the particle just encircles the axis of the field as shown  in Fig. \ref{four_tr}, segments (a) and (c). 
\begin{figure}
  \centerline{\includegraphics[width=0.8\columnwidth]{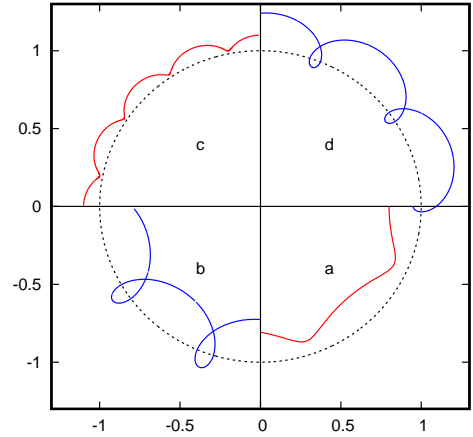}}
  \caption{Trajectories of a charged particle in the potential wells shown in Figs \ref{four1} and \ref{four2}. Segments (a) and (b) correspond to potential of shape (b) in Fig. \ref{four1} or of shape (a) in Fig. \ref{four2}. The  segments (c) and (d) correspond to the  potential of shape (a) in Fig. \ref{four1}. Low-energy particles do not change their azimuthal velocity (segments (a) and (c)).}
\label{four_tr}
\end{figure}
 With increasing energy, the amplitude of  radial oscillations increases and at a certain energy the particle reaches the circle of radius $\rho=1$.  Consequently, as $\rho$ becomes greater than unity, the azimuthal velocity of the particle begins to change its sign and the trajectory acquires a loop-like shape like in segments (b) and (d) of Fig. \ref{four_tr}.  Similar behaviour show the particle trajectories in the magnetic field with superimposed gravitational field \citep{Bellan2007} or  in the combination of a uniform axial magnetic field and parabolic electrostatic potential \citep{Bellan2016_PP}.

 The trajectories are obtained by   numerical integration of the exact equations of motion. The guiding centre approximation is  applicable under nearly  the same conditions as in the dipole magnetic field. This approximation can be  used when the trajectory curvature 
\begin{equation}
r_c=\frac{c^2\gamma m}{eH^2}\left| \bm\beta\times \bm H+\bm E\right |
\end{equation}
is much smaller  than the size of the area within the particle moves \citep{Alfven1950}. The  induced electric field is of order $E\sim a^2\omega H/cr$. But $a\omega$ is the speed of the points on the equator of a rotating celestial body. Therefore,  if the particle moves much faster than the points at the equator (which is usually the case), then the second term in the above equation is much less than the first one. Hence, the  induced electric field has little effect on the trajectory curvature.
}
%
\section{Stationary points of effective potential energy}\label{sec3}
%
%
The stationary points of effective potential energy $V(\rho,\theta)$ are defined be equations $\partial V/\partial \rho=0$ and $\partial V/\partial \theta=0$.
 Equation $\partial V/\partial \rho=0$ gives
\begin{multline}\label{Rm}
\rho\,(\rho-\epsilon\sin^2\theta)(2\epsilon\sin^2\theta-\rho)\\
=\epsilon\Phi\sin\theta (3\sin^2\theta-2)\sqrt{\rho^4\sin^2\theta+P\left(\rho-\epsilon\sin^2\theta\right)^{2}},
\end{multline}
while  $\partial V/\partial \theta$ is equal to zero in the equatorial plane $\theta=\pi/2$  and on the line defined by equation 
\begin{equation}\label{Tm}
\rho(\rho^2-\sin^4\theta)=2\epsilon\Phi\sin^3\theta\sqrt{\rho^4\sin^2\theta+P\left(\rho-\epsilon\sin^2\theta\right)^{2}}.
\end{equation}
In the equatorial plane $\theta = \pi/2$ the coordinates of the stationary points can be found directly from  equation (\ref {Rm}), which in this case takes the form
\begin{equation}\label{Rm1}
\rho(\rho-\epsilon)(2\epsilon-\rho)=\epsilon\Phi\sqrt{\rho^4+P(\rho-\epsilon)^{2}}.
\end{equation}
 We assume that  $\theta\neq 0,\pi$, because the potential (\ref{V}) is infinitely large in these directions, unless $M=0$. Hence, the axis $\theta= 0,\pi$ is inaccessible for the particles with non zero $M$. 
Particles with $M=0$ will be considered later.

\begin{figure}
\centerline{\includegraphics[width=\columnwidth]{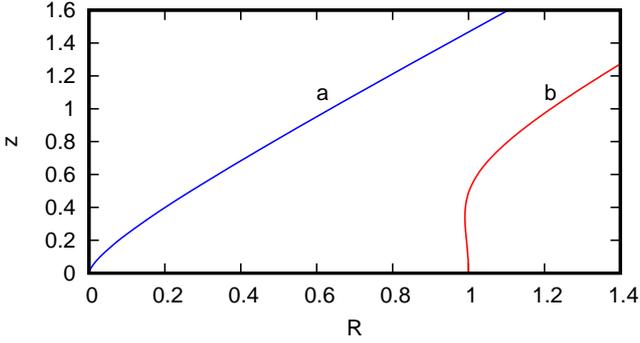}}
\caption{Lines of stationary points of the  potential $V(\rho,\theta)$ for $\epsilon=-1$ (a) and $\epsilon=1$ (b).}
\label{extrA}
\end{figure}

Equations (\ref {Rm}) - (\ref {Rm1}) are algebraic equations of the degree six and higher, and can not be solved in term of radicals\footnote{Actually, the equation of contour lines of  potential  (\ref{V}) can be represented in analytical form, since by assigning a specific value for $V$, we obtain a cubic equation  with respect to $\sin^2\theta$. However, this approach does not much facilitate  analytical study of the potential $V$.}. Nevertheless, some conclusions about the stationary points can be made in a general form. Eliminating the square root from  equations (\ref {Rm}) and (\ref {Tm}), we obtain
\begin{equation}\label{RT}
\rho=\epsilon\sin^2\theta\frac{\sin^2\theta+2}{5\sin^2\theta-2}.
\end{equation}
This equation defines two lines plotted in Fig.~\ref{extrA} in cylindrical coordinates $R=\rho\sin\theta$ and $z=\rho\cos\theta$.
Remarkable, that all the off-equatorial stationary points lie on one of these lines regardless of the values of $P$ and $\Phi$.
If $\epsilon=1$, the line (\ref{RT}) lies in the domain $\rho> 1, \, \sin\theta> \sqrt{2/5}$,  while in the case $\epsilon=-1$ it lies  in the region  $\rho>0, \, \sin\theta<\sqrt{2/5}$. It will be shown below, that there are no more than two symmetrical off-equatorial stationary points in each case, and these are the saddle points.

\begin{figure}
\centerline{\includegraphics[width=\columnwidth]{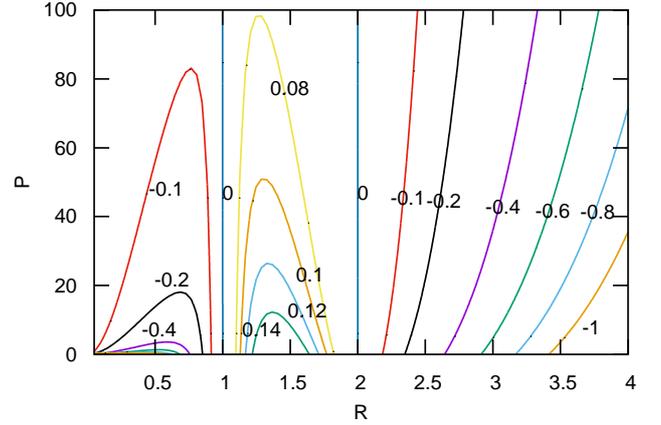}}
\caption{Coordinate $R$ of the stationary points in equatorial plane as a function of  $P$ and $\Phi$ for the case $\epsilon=1$. Labels on the lines
  $\Phi = {\rm const}$ indicate the values of $\Phi$. There are no  stationary points in the interval $1 < R <2 $ for  $\Phi>3-2\sqrt{2}$.}
\label{izo-eqv1}
\end{figure}

Coordinates of the stationary points in the equatorial plane are determined by  equation (\ref{Rm1}). To calculate  the number of roots of this equation we express $P$ as a function of $R$ ($\rho=R$ in the equatorial plane)
\begin{equation}
P( R)=\frac{ R^2(2\epsilon- R)^2}{\Phi^2}-\frac{ R^4}{( R-\epsilon)^2}.
\end{equation}
The two extrema of this function in domain $P>0$ are defined by equation
\begin{equation}\label{or}
\sqrt{2}( R-\epsilon)^2= R|\Phi|.
\end{equation}
 We start with the case $\epsilon = 1$. One can see from equation (\ref{Rm1}) that $\Phi$ is positive on  interval $1 < R <2$ and negative or zero for other values of $ R$. Having this in mind, we deduce from equations  (\ref{or}) that $P( R)$ has one maximum in each of intervals $0 < R <1 $ and  $1 < R <2 $ and no extrema in the domain $ R>2$. However, the maximum value of $P( R)$ in the interval  $1 < R <2$ turns out to be negative if $\Phi>3-2\sqrt{2}$. Hence, there are no stationary points in this interval for $\Phi>3-2\sqrt{2}$, since $P$ is positive by definition.
Plots of the positive part of function $P( R)$  for a set of  values of $\Phi$ are shown in Fig.~\ref{izo-eqv1}. 
The diagram represents positions of the stationary points versus $\Phi$ and $P$ as a map of lines $\Phi = {\rm const}$ for $\epsilon = 1$. For example, for $\Phi = -0.1$ and $P = 40$, the stationary points lie at the points $R_1\approx 0.4, \, R_2\approx 0.9$ and $R_3\approx 2.3$. 

\begin{figure}
\centerline{\includegraphics[width=\columnwidth]{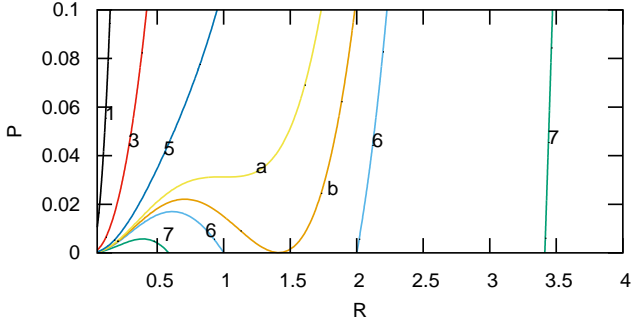}}
\caption{Coordinate $R$ of the stationary points in equatorial plane as a function of  $P$ and $\Phi$ for the case $\epsilon=-1$. Labels on the lines
  $\Phi = {\rm const}$ indicate the value of $\Phi$. The lines labelled by `a' and `b' correspond to $\Phi=4\sqrt{2}$ and  $\Phi=2\sqrt{2}+3$ respectively.}
\label{izo-eqv2}
\end{figure}

Applying the same analyses to the case $\epsilon=-1$, we find that the function $P(R)$ increases monotonically with $R$ if $\Phi<4\sqrt{2}$, and has extrema if $\Phi>4\sqrt{2}$. If $\Phi> 2\sqrt{2}+3$, the  function $P(R)$ has two branches on different sides of the point $R=\sqrt{2}$. These peculiarities are seen in Fig.~\ref {izo-eqv2}.
Hence, in the case $\epsilon=-1$, the potential $V$ can have one or three stationary points in the equatorial plane. In the last case two stationary points are positioned in the region $R<\sqrt{2}$, and one point at $R>\sqrt{2}$. There are no stationary points if $\Phi$ is negative. Corresponding examples are given in the section~\ref{examp}.
%

The nature of  stationary points (whether minimum or maximum) in the equatorial plane can be determined by use of the asymptotic of the potential $V (R)$.
If $\rho \to 0$, the potential tends to infinity if $\epsilon \Phi> 0$ and to minus infinity if $\epsilon \Phi < 0$. Hence, if $\epsilon \Phi> 0$, the stationary point with the smallest $R$ is a minimum,  the next point is a maximum, etc. Typical graphs of $V (R)$ for four possible combinations of signs of $\epsilon$ and $\Phi$ are shown in Fig.~\ref {four1} and Fig.~\ref {four2}.

Positions of the stationary points outside the equatorial plane are determined by the system of equations (\ref{Rm}) and (\ref{Tm}), or  equation (\ref{RT}) and one of the equations (\ref{Rm}) or (\ref{Tm}). The left-hand side of  equation (\ref{Tm}) is always positive. For $\epsilon = 1$ it is obvious, because $\rho> 1$  according to  equation (\ref{RT}).  It follows from equation (\ref {RT}) that in the case $\epsilon = -1$,  the value of $\rho-\sin ^ 2\theta$ is positive: $\rho-\sin ^ 2\theta = 6\sin ^ 4 (2 -5\sin ^ 2\theta)> 0$. Consequently, the right-hand side of  equation (\ref {Tm}) is also positive. Thus, the stationary points outside the equatorial plane exist only if $\epsilon\Phi> 0$. 

The potential (\ref{V}) has a singularity at the origin. If $\rho\to 0$, only the last term remains in equation (\ref{V}). It tends to $\pm\infty$ depending on the sign of  expression $e\omega (1-3\cos ^ 2\theta )$. For example, in case of $e\omega> 0$,
$V\to-\infty$ in the region of the angles $\sin\theta <1 /\sqrt {3}$ and $V\to\infty$ in the domain $\sin\theta\geq 1 /\sqrt {3}$. 

\begin{figure}
\centerline{\includegraphics[width=\columnwidth]{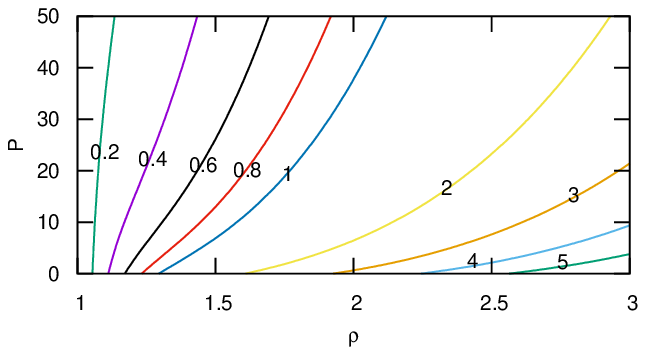}}
\caption{Coordinate $\rho$ of the off-equatorial stationary points  for $\epsilon = 1$.
Labels on the lines  $\Phi = {\rm const}$ indicate the value of $\Phi$.}
\label{izo1}
\end{figure}

\begin{figure}
\centerline{\includegraphics[width=\columnwidth]{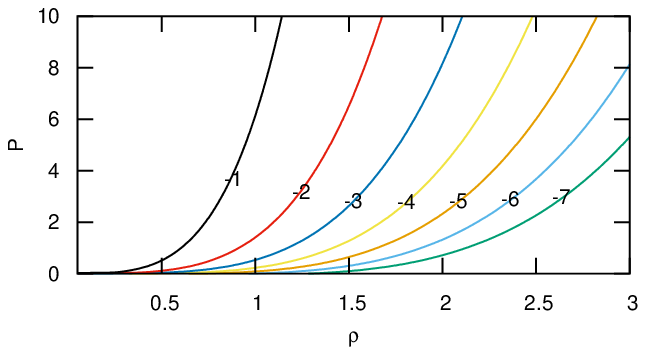}}
\caption{Coordinate $\rho$ of the off-equatorial stationary points  for $\epsilon = -1 $.
Labels on the lines  $\Phi = {\rm const}$ indicate the value of $\Phi$.}
\label{izo2}
\end{figure}

In order to find the coordinates of the stationary points outside the equatorial plane, we express $\sin ^ 2\theta $ from  equation (\ref{RT})
\begin{equation}\label{u}
\sin^2\theta=\frac 12\left[5\epsilon \rho-2-\epsilon\sqrt{(5\epsilon \rho-2)^2-8\epsilon \rho} \right]
\end{equation}
and substitute it into  equation (\ref {Tm}). The result of this substitution looks rather cumbersome and we don't produce it here. The numerical solution of the equation obtained is shown in Fig.~\ref {izo1} and Fig.~\ref {izo2}.

As shown above, the off-equatorial stationary points exist only if $\epsilon\Phi> 0$. It follows from the figures that there is only one value of coordinate $\rho$ for each given values of $P$ and $\Phi$.
If $\epsilon = 1$, the stationary point lies on the curve (b) in Fig.~\ref{extrA} and has the radial coordinate $\rho> 1$. In the case  $\epsilon = -1$
the stationary point is positioned on  curve (a) in Fig.~\ref {extrA}. Therefore, all curves in  Fig.~\ref {izo2} start from the origin.

Knowing the radial coordinate of the stationary point, we can use  equation (\ref{u}) to find the coordinate $\theta$. Since $\sin\theta $ is symmetric relative to the equatorial plane, there are two symmetric stationary points of the potential $V(\rho,\theta)$. In  section \ref{examp} we demonstrate that these points are the saddle points.

{
All what is said above refers to particles of arbitrary energy.  For a non-relativistic particle (low initial kinetic energy and relatively weak electric field), one has to put $P\ll 1$. For example, the lines $\Phi = {\rm const}$ in Figs \ref{izo-eqv1} -- \ref{izo1} become just vertical lines, except the line (b) in Fig. \ref{izo-eqv2}. Hence, the positions of the stationary points for a non-relativistic particle do not generally depend on $P$. As to the case $\epsilon = -1 $ (Fig.~\ref{izo2}), a non-relativistic particle has no off-equatorial stationary points.
}
%
\section{Particles with zero angular momentum}\label{m_zero} 
%
So far $M$ is taken to be non zero.
Let us consider particles with  zero angular momentum~$M$. We cannot use expression  (\ref{V}) in this case, because the scale factor $\Gamma$ is equal to zero. Putting $M=0$ in equation (\ref{U}), and using new scale factor for the radial coordinate $\rho'=r\sqrt{mc^2/|e|\mu}$, we rewrite the potential $V$ in the form
\begin{equation}\label{V1}
V=\sqrt{1+\frac{\sin^2\theta}{\rho'^4}}-1+\frac{\Phi'}{3\rho'^3}(1-3\cos^2\theta), 
\end{equation}
where
\begin{equation}
\Phi'=a^2\omega\frac{e}{e_0}\sqrt{\frac{m}{e_0\mu}},\quad e_0=|e|.
\end{equation}
Note that this time the potential $V$ is finite on the axis $\theta=0,\pi$ except the origin $z=0$. 

It follows from equation (\ref{p3}) that the initial condition $M=0$ implies that $\mathring\psi$ does not change sign during the particle motion, i.e. the particle is just revolving around the axis~$z$. 

Let us find the stationary points. Equation $\partial V/\partial \rho'=0$ results in
\begin{equation}\label{st1}
2\rho'\sin^2\theta -\Phi'(1-3\cos^2\theta)\sqrt{\rho'^4+\sin^2\theta}=0,
\end{equation}
and  equation $\partial V/\partial \theta=0$  leads to two equations: either
\begin{equation}\label{st2}
\rho'+2\Phi'\sqrt{\rho'^4+\sin^2\theta}=0,
\end{equation}
or
\begin{equation}
\sin 2\theta=0.
\end{equation}
It is easy to prove that equations (\ref{st1}) and (\ref{st2}) have not common roots. Hence, all stationary points lie in the equatorial plane $\theta=\pi/2$ with its radial coordinate defined by equation  (\ref{st1}) at $\theta=\pi/2$. This equation has solution only if $\Phi'\leq 0$, which reads
\begin{equation}\label{roots}
\rho'_{1,2}=-\frac{1}{\Phi'}\left(2\pm\sqrt{4-\Phi'^4}\right)^{1/2},\quad -\sqrt{2}\leq\Phi'\leq 0.
\end{equation}
The root with the plus sign corresponds to the minimum, while the second root specifies the maximum of potential $V(\rho')$. The values of potential at these points are
\begin{equation}\label{Vm}
V_{1,2}=-\frac{4\pm \sqrt{4-\Phi'^4}}{3\Phi'\rho'_{1,2}}-1.
\end{equation}

\begin{figure}
\centerline{\includegraphics[width=\columnwidth]{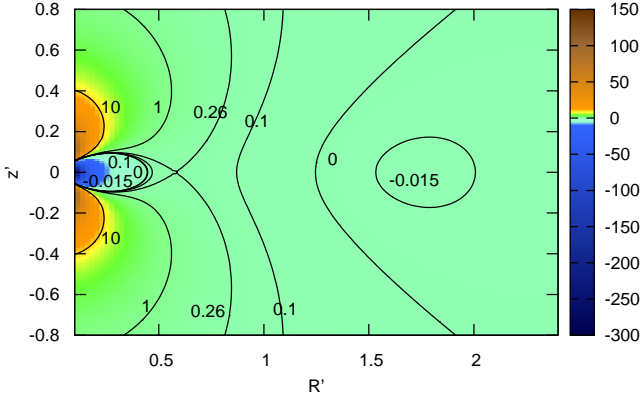}} 
\caption{Contour lines of the potential $V$ for $M=0,\,\Phi'=-1.1$.}
\label{N=0}
\end{figure}

An example of  topography of potential $V$  is shown in Fig.~\ref{N=0} in cylindrical coordinates $R'=\rho'\sin\theta$ and $z'=\rho'\cos\theta$.
Inspection of the contour map shows that there is an infinitely deep potential well at the centre and a side well with its bottom at $z=0,\, R\approx 1.7$. The  wells are connected by a pass at $z=0,\, R\approx 0.6$. Coordinate of the pass and that of the minimum of potential energy is defined by eq. (\ref{roots}). The height of the pass ($V_2\approx 0.26$) and the depth of the side well are given by eq. (\ref{Vm}). Hence, the particles with energies $({\cal E}-mc^2)/mc^2<V_2$ are confined to the central well, while the particles with energies $V_1<({\cal E}-mc^2)/mc^2<0$ must remain in the side well, restricted by the contour line $V=0$.
\begin{figure}
\centerline{\includegraphics[width=0.48\columnwidth]{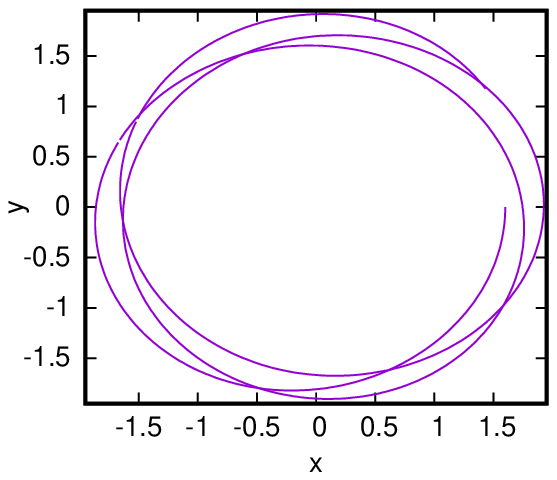}\hfill
\includegraphics[width=0.48\columnwidth]{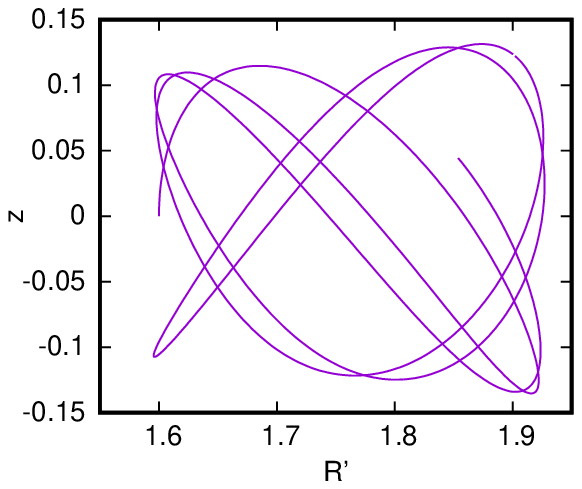}} 
\caption{Trajectory of a particle with $M=0$ and normalized energy ${(\cal E}-mc2)/mc^2=-0.015$ in the field with $\Phi'=-1.1$. Projection of the trajectory in the x-y plane (left) and dependence of $z$-coordinate on $R'$  (right).} 
\label{trN=0}
\end{figure}
{
Typical trajectory in this side well is shown in Fig. \ref{trN=0}.

A particle in the central well, having energy below that of the pass, spirals down to the origin, no matter what the initial velocity is (provided however, that $M=0$). The particle trajectories in this case are similar to the trajectories in the gravitational field combined with the electric and magnetic fields studied by \citet{Bellan2016}. The fall of the particle to the centre in our case is due to relatively steep gradient of the effective potential energy: it tends to $-\infty$ as $-1/\rho'^3$, while the ``centrifugal potential'' decreases only  as  $-1/\rho'^2$ \citep[\S 14]{Landau_I}.
}
\section{Examples of  possible potential energy configurations}\label{examp}
\begin{figure}
\centerline{\includegraphics[width=\columnwidth]{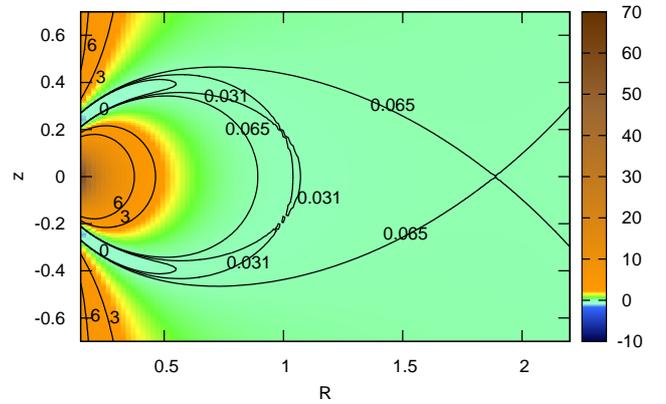}}
\caption{A contour map 
for $\epsilon=1,\,\Phi=0.1$. }
\label{fig7}
\end{figure}

\begin{figure}
\centerline{\includegraphics[width=\columnwidth]{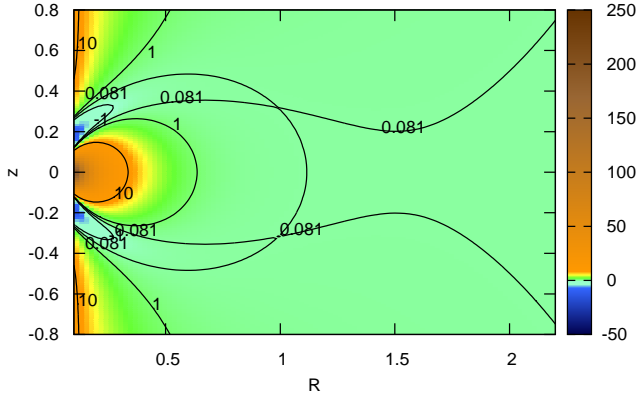}}
\caption{A contour map of the potential $V$ 
for $\epsilon=1,\,\Phi=0.3$. }
\label{fig8}
\end{figure}

We will now proceed to show examples of the potential energy topography for each  of the possible combinations of  signs of $\Phi$ and  $\epsilon$.
Al figures in this section are plotted for $P=2$. A contour map of the potential $V$ for $\epsilon=1,\,\Phi>0$ is shown in Fig.~\ref{fig7}.
According to the inequality (\ref{ener1}), the contour line $V = ({\cal E} -mc ^ 2) / mc ^ 2 $ limits the region of motion of a particle with energy $ {\cal E}$. 
A region allowed for a charged particle motion will be referred to as inner or closed region, if it does not include infinitely distant point.
 Conversely, an allowed region that extends to infinity will be called external or open one.
 In the case shown in Fig.~\ref{fig7} 
we see an internal potential valley bounded by contour lines  $V\approx 0.065 $, which connects to the outer open valley through the pass at the point $z = 0, \, R\approx 1.9$. All other paths leading from the inner valley to infinity and not going through the pass encounter larger values of V.  In the inner valley, one can distinguish three regions -- a narrow potential well in the equatorial plane, whose bottom lies at the point $z = 0, \, R\approx 1.03$, and two potential valleys sloping down towards the polar regions. The circumpolar valleys are connected with the equatorial well by  two saddle points at $z\approx\pm 0.18, \,R\approx 1$. Potential $V$ in the equatorial plane varies according to Fig.~\ref{four1} (a). Both stationary points, the bottom of the potential well and the pass, lie in the interval $1 <R<2$ in compliance  with  Fig.~\ref{izo-eqv1}. With increasing $\Phi$ or $P$, these stationary points approach each other as can be seen from the diagram in Fig.~\ref{izo-eqv1}, and for some $\Phi $ and $ P $ merge into one point.
For greater values of $\Phi$ or $P$, there are no stationary points in the equatorial plane: the function $V (R)$ decreases monotonically with the growth of $R$.
An example of this kind of geometry is given in Fig.~ \ref{fig8}.
Unlike the previous example, the off-equatorial saddle points 
connect the sub-polar valleys with the outer open valley.

\begin{figure}
\centerline{\includegraphics[width=\columnwidth]{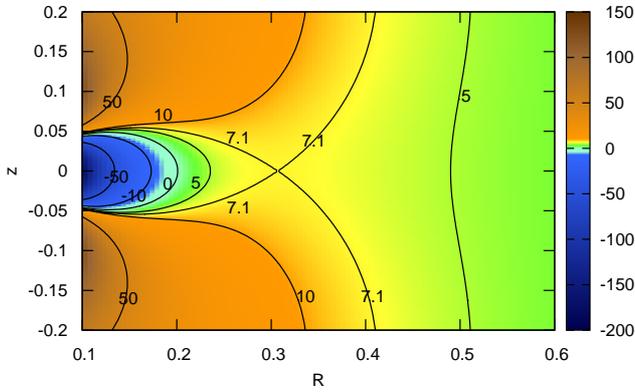}}
\caption{A contour map for  $\epsilon=-1,\,\Phi=1$. }
\label{fig9}
\end{figure}

Let us consider the case $\epsilon = -1, \,\Phi> 0$ (corresponding to the figure \ref{four2}(a)). According to discussion in section \ref{sec3}, the potential energy in this case has  one or three stationary point in the equatorial plane. A contour map for specific values of $\Phi$ and $P$ is exemplified  in Fig.~\ref {fig9}.
The potential energy contains an infinitely deep potential well in the centre of the field, which is connected with the outer region by a pass in the equatorial plane. The saddle point lies at the level $V \approx 7.1$. With increasing $P$ or / and $\Phi$, the pass moves away from the centre of the field, as it follows from Fig.~\ref {izo-eqv2}.

\begin{figure}
\centerline{\includegraphics[width=\columnwidth]{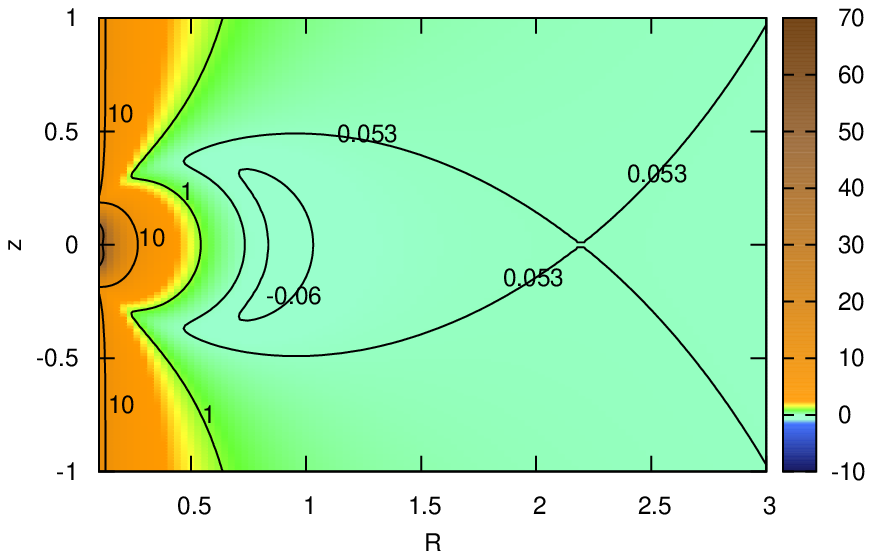}}
\caption{A contour map for  $\epsilon=1,\,\Phi=-0.2$.}
\label{fig10n}
\end{figure}

In the case $\epsilon = 1, \, \Phi <0$ (Fig.~\ref {four1}(b)), there are one or three stationary points in the equatorial plane. There are no stationary points outside the equatorial plane, since $\epsilon \Phi$ is negative. Geometry of the  potential with three stationary points is shown in Fig.~\ref{fig10n}. It has
\begin{figure}
\centerline{\includegraphics[width=\columnwidth]{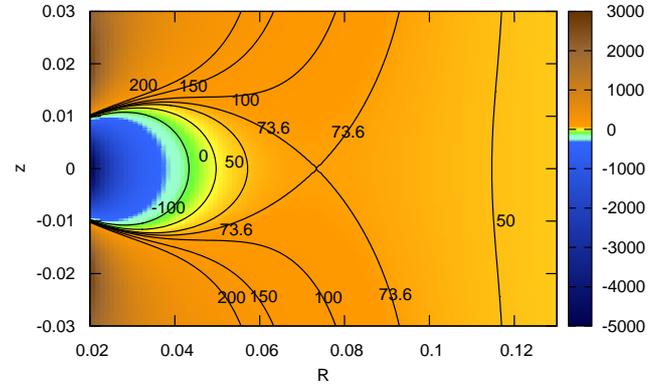}}
\caption{Central part of the contour map of Fig.~\ref{fig10n}.} 
\label{fig10c}
\end{figure}
 a potential valley 
 bounded by the contour  line  $V\approx 0.053$. The pass that  connect this potential valley to the outer region has  coordinates $R\approx 2.2, \, z = 0$. With the variation of $P$ and $\Phi$, the stationary points move in accordance with the diagram in Fig.~\ref {izo-eqv1}. Fine details in the centre of the field are shown in Fig.~\ref {fig10c}. For  sufficiently large $P$ or large negative $\Phi$, the pass illustrated in Fig.~\ref {fig10c}  drops down and disappears. Then, only infinitely deep potential well in the centre remains with a pass in the region $R> 2$.
 
\begin{figure}
\centerline{\includegraphics[width=\columnwidth]{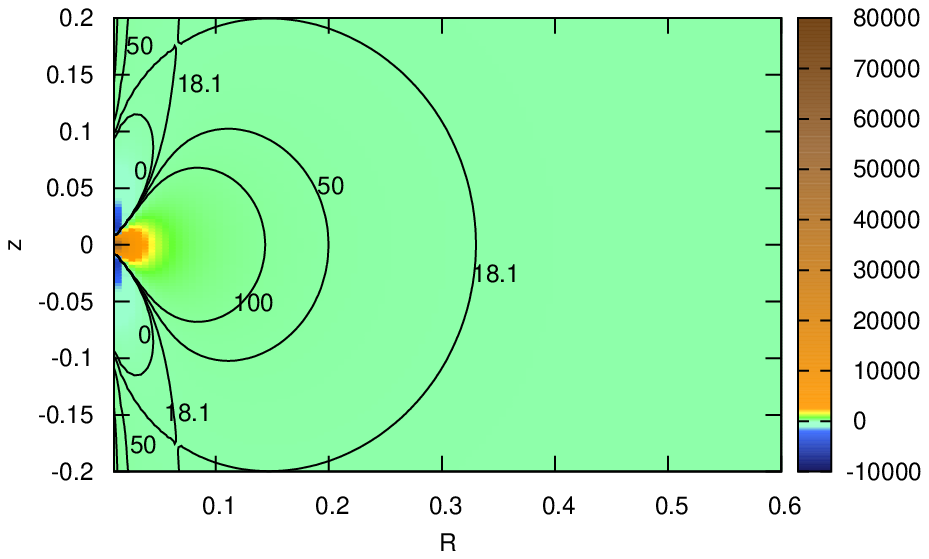}}
\caption{A contour map 
for  $\epsilon=-1,\,\Phi=-0.2$.}
\label{fig11n}
\end{figure}

Finally, we illustrate the case, corresponding to Fig.~\ref {four2}(b): $\epsilon = -1, \, \Phi <0$. The potential topography for  $\Phi=-0.2$ is shown in Fig.~\ref {fig11n}.
There are no stationary points at equatorial plane.
Two symmetrical passes with coordinates $\rho \approx 0.07, \, z = \pm 0.17$ connect the polar valleys  bounded by the contour line $V \approx 18.1$ with a broad open valley.

  It is easy to verify that the coordinates of the off-equatorial stationary points shown in the above figures  satisfy  equation (\ref {RT}).
  
\section{Results}
The presence of a quadrupole electric field significantly changes the geometry of the effective potential energy. A purely dipole magnetic field can capture particles only with a positive value of $eM$ (where $M$ is the initial canonical angular momentum of the particle with respect to  $z$ axis). As to the field of a coaxial rotator, there are trapping regions for particles with both positive and negative values of this product. The influence of the electric field is determined by the direction of rotation of the celestial body and the sign of the particle charge . 

	If $e\omega> 0 $, the quadrupole electric field increases the particle potential (\ref{V}) in the equatorial region ($\sin\theta> 1 / \sqrt{3} $) and decreases it in the near-polar regions ($\sin\theta <1 / \sqrt{3} $). As a result, for relatively small values of the electric field, the predicted by St{\o}rmer potential valley with its bottom along the magnetic field line $\rho = \sin ^ 2 \theta$ breaks up into a toroidal trapping region in the equatorial plane and two potential valleys in the near-polar regions (Fig.~\ref{fig7}). With the growth of the electric field, the equatorial trapping region vanishes and only the sub-polar valleys remain. In this case, the pass to the open region, which in the St{\o}rmer potential lies  in the equatorial plane, splits into two off-equatorial passes  (Fig.~\ref{fig8} and Fig.~\ref{fig11n}).
	
In the case of $e\omega <0 $, all the trapping regions  are centred at the equatorial plane and are connected to each other or to the open region by passes that lie in this plane. One of these areas is an infinitely deep potential well at the origin (Figs.~\ref{fig9} -- \ref{fig10c}). If $eM> 0$, then a closed toroidal trapping region in the equatorial plane  also exists (Fig.~\ref{fig10n}).

The boundaries of the trapping regions  are defined by  condition ${\cal E} \ge U$. However, it does not mean that the particles having energy ${\cal E}$ fill the entire thus defined region. In particular, 
 the first adiabatic invariant associated with the conservation of the magnetic flux, which the particle covers in one revolution around the magnetic field line, implies additional restriction on the motion  within the allowed region. Conservation of this flux leads to  reflection of the particles from the regions with  stronger magnetic field.
In our case, the potential valleys in the polar regions reach formally the centre of the field. However, the particles having a sufficiently large pitch angle can be reflected by a `magnetic mirror', not reaching the surface (or the dense layers of  atmosphere) of the celestial body. If the electric field tangential to the magnetic field line is not zero, the position of the `magnetic mirror' is shifted relative to its position in a pure magnetic field, by a distance depending on the particle pitch angle and  relation between the tangential electric field and the magnetic field gradient \citep{Alfven1963}. Thus, in the case of $e\omega> 0$ and  not very strong electric field, the polar valleys are actually confined by the `magnetic mirrors'  from the side of the celestial body. This makes possible the existence of the closed toroidal off-equatorial radiation belts.
\section{Applicability of the model to real celestial bodies}

In this section, we  discuss the issue of how the idealized model of a rotating uniformly magnetized sphere with a vacuum magnetosphere is applicable to real celestial bodies. To this end, we calculate the model parameters for four celestial bodies: Earth, Jupiter, brown dwarf  and a neutron star. The Earth and the neutron star are chosen as examples of bodies with a very weak and extremely strong induced electric field. Jupiter and a  brown dwarf are real bodies to which the model is believed to be applicable because they are highly magnetized and relatively fast rotating objects. The  brown dwarfs  exhibit strong magnetic fields and it seems to be auroras on brown dwarfs  \citep{Berdyugina2017,Pineda2017, Kao2016}.

In order to estimate the extent to which the particle dynamics is affected by  the quadrupole electric field, we  can compare the electric field potential (\ref{w}) in the vicinity of the body with the particle rest energy. This can be made by use of a dimensionless parameter
\begin{equation}\label{eta}
\eta = \frac{e \mu \omega}{3 a m c^3} \sim \frac{e}{ mc ^ 2} \varphi\,(a). 
\end{equation}
If $\eta \gtrsim 1$, then the electric field can accelerate the particles to relativistic energies. 
{
The values of $\eta$ for the four celestial bodies are presented in Table \ref{tab1}. In case of a brown dwarf and a neutron star, some typical values  of $a$,  $\omega$ and $\mu$ are taken.
One can see that the influence of the quadrupole electric field on the Earth's magnetosphere can be neglected, especially as the Earth`s dipole  field  is highly  distorted by the solar wind. However, the  electric field can substantially change the structure of the radiation belts in the vicinity of the three other celestial bodies. According to equation (\ref{p0}),  variations of the potential on the  particle trajectory are equal to  variations of $mc^2\mathring t$. Since $\mathring t = \gamma$, the value of $\eta$ defines the variations of  the particle relativistic factor along the trajectory: $\Delta\gamma\sim \eta$.

The vacuum magnetosphere model  is applicable to the case of   not very strong electric field. A sufficiently strong electric field can extract charged particles from the surface of the celestial body. The magnetosphere thus obtains an unlimited source of charged particles and will be filled with a plasma that shorts out the longitudinal electric field. The applicability limits of the model in this sense can be estimated by comparing the magnitude of the induced electric field with the field at which the field electron emission from the solid surface begins. A typical value of the electric field at which the  electron emission from a cold surface starts is about $10^9$~V/m $\approx 3\times 10^4$ CGS units. The  electric field 
$E_i$ on the surface of a celestial body can be estimated as 
\begin{equation}\label{Ei}
E_i=|\bm\nabla\phi(r)|\sim \frac{\mu\omega}{ca^2}. 
\end{equation}
As can be seen from the table, only the field of a neutron star exceeds the threshold of cold emission. 
 However, even for the neutron stars our  model can be useful for understanding   the initial stage of filling the magnetosphere with plasmas.
\begin{table}
	\centering
	\caption{Estimation of the model parameters for some celestial bodies. Numbers in parentheses indicate the number of equation which defines the parameter. Parameters that depend on the charge and mass of the particle are calculated for an electron. CGS system of units is used.}
	\label{tab1}
	\begin{tabular}{lllll} 
		\hline
		Parameter  & Eath & Jupiter &Brown& Neutron\\
		& & &  dwarf& star\\
		\hline
		$a$ & 6$\times 10^8$ & 7$\times 10^9$ & 7$\times 10^9$ & $10^6$\\
		$\omega$ & 7$\times 10^{-5}$ & 2$\times 10^{-4}$ & 6$\times 10^{-4}$ & 10\\
		$\mu$ & 8$\times 10^{25}$ & 2$\times 10^{30}$ & $10^{32}$ & $10^{30}$\\
		$\eta$ (\ref{eta})& 6$\times 10^{-2}$ & 2$\times 10^{2}$ & 5$\times10^{4}$ & 7$\times 10^{10}$\\
		$E_{i}$ (\ref{Ei})& 5$\times 10^{-7}$ & 2$\times 10^{-4}$ & 4$\times 10^{-2}$  & 3$\times 10^{8}$\\
		\hline
	\end{tabular}
\end{table}
Of course, the  magnetosphere of  real objects is filled with more or less dense  plasmas. Unfortunately, the  models with a plasma-filled magnetosphere do not have an analytical solution. Most of the latest work in this area are devoted to numerical modelling of plasma dynamics \citep{Cerutti2017}. However, the  numerical methods  give no insight into general properties of a solution, they provide only specific answers to the  problem at specific values of the parameters. An idealized vacuum model allows analysing the  morphology of the potential in a wide range of parameter values and can give a hint what can be expected in a certain range of these  parameters and what types of orbits  can occur.

The position and dimensions of the radiation belts are determined by the scale factor $ |\Gamma|$. In the absence of an electric field, the bottom of the potential valley lies on the line $\rho_ {min} = \sin^2 \theta$. The actual position of the minimum of the potential energy depends on the initial value of the canonical angular momentum $M$:
\begin{equation}\label{rmin}
r_{min}=\left |\frac{e\mu}{Mc}\right| \sin^2\theta.
\end{equation}
Accordingly, particles with a large value of $ M $ are moving closer to the surface of the celestial body, while particles with a small $ M $ occupy a more remote region. Other conditions being equals, particles with a larger mass have a larger angular momentum, so the radiation belt of protons is closer to the surface  than the electron belt. On the inside, the radiation belts are 	constrained by the surface of the celestial body or by  a fairly dense atmosphere. On the outside, the radiation belts, in principle, are not limited. The actual dimensions of the radiation belts depend on the source of the particles in these belts, namely, on what values of $M$ the source supplies the particles. 
For example, the main sources of charged particles in the Earth's radiation belts are  the plasma streaming from the Sun and the ``albedo''  neutrons.
 All that has been said above remains valid in the presence of a quadrupole electric field. However, the electric field shifts the position of the potential valley in the direction of the vector $e\bm E$ and transforms the shape of the effective potential as discussed in Section \ref{qual}.

Another issue that should be discussed is the radiation friction. If the electric and magnetic fields are relatively  large, then the high-energy particles can be trapped. The relativistic  particles undergo intensive radiative friction, leading to particle energy losses. Though, the particle motion with respect to the radiation reaction is not a subject of this paper, the effective potential energy is nevertheless, a powerful instrument of qualitative analysis of the particle behaviour in this case. Because the structure of the potential energy is defined solely by the field and initial conditions and does not depend on the particle motion. It seems obvious that if a particle loses its energy due to radiation, it progressively passes to lower energy levels continuing its motion around other magnetic field lines. Hence, the boundary of the allowed region changes with time, shifting the particle downhill the potential profile.
}

\section*{Acknowledgement}
This work was supported by a grant of the Ministry of Education and Science of the Russian Federation under project No. 3.1386.2017





\begin{thebibliography}{}
\makeatletter
\relax
\def\mn@urlcharsother{\let\do\@makeother \do\$\do\&\do\#\do\^\do\_\do\%\do\~}
\def\mn@doi{\begingroup\mn@urlcharsother \@ifnextchar [ {\mn@doi@}
  {\mn@doi@[]}}
\def\mn@doi@[#1]#2{\def\@tempa{#1}\ifx\@tempa\@empty \href
  {http://dx.doi.org/#2} {doi:#2}\else \href {http://dx.doi.org/#2} {#1}\fi
  \endgroup}
\def\mn@eprint#1#2{\mn@eprint@#1:#2::\@nil}
\def\mn@eprint@arXiv#1{\href {http://arxiv.org/abs/#1} {{\tt arXiv:#1}}}
\def\mn@eprint@dblp#1{\href {http://dblp.uni-trier.de/rec/bibtex/#1.xml}
  {dblp:#1}}
\def\mn@eprint@#1:#2:#3:#4\@nil{\def\@tempa {#1}\def\@tempb {#2}\def\@tempc
  {#3}\ifx \@tempc \@empty \let \@tempc \@tempb \let \@tempb \@tempa \fi \ifx
  \@tempb \@empty \def\@tempb {arXiv}\fi \@ifundefined
  {mn@eprint@\@tempb}{\@tempb:\@tempc}{\expandafter \expandafter \csname
  mn@eprint@\@tempb\endcsname \expandafter{\@tempc}}}

\bibitem[\protect\citeauthoryear{Alfven}{Alfven}{1950}]{Alfven1950}
Alfven H.,  1950, Cosmical Electrodynamics.
Oxford, Oxford Clarendon Press

\bibitem[\protect\citeauthoryear{Alfv{\'e}n \& F{\"a}lthammar}{Alfv{\'e}n \&
  F{\"a}lthammar}{1963}]{Alfven1963}
Alfv{\'e}n H.,  F{\"a}lthammar C.,  1963, Cosmical electrodynamics: fundamental
  principles.
International series of monographs on physics, Oxford, Oxford Clarendon Press

\bibitem[\protect\citeauthoryear{{Asseo}, {Beaufils}  \& {Pellat}}{{Asseo}
  et~al.}{1984}]{Asseo1984MNRAS}
{Asseo} E.,  {Beaufils} D.,   {Pellat} R.,  1984, \mn@doi [\mnras]
  {10.1093/mnras/209.2.285}, \href
  {http://cdsads.u-strasbg.fr/abs/1984MNRAS.209..285A} {209, 285}

\bibitem[\protect\citeauthoryear{Bellan}{Bellan}{2007}]{Bellan2007}
Bellan P.~M.,  2007, \mn@doi [Phys. Plasmas] {10.1063/1.2815791͔}, 14, 122901

\bibitem[\protect\citeauthoryear{Bellan}{Bellan}{2016a}]{Bellan2016_PP}
Bellan P.~M.,  2016a, \mn@doi [J. Plasma Phys] {10.1017/S0022377816000064}, 82,
  615820101

\bibitem[\protect\citeauthoryear{Bellan}{Bellan}{2016b}]{Bellan2016}
Bellan P.~M.,  2016b, \mn@doi [MNRAS] {10.1093/mnras/stw562}, 458, 4400

\bibitem[\protect\citeauthoryear{Berdyugina, Harrington, Kuzmychov, Kuhn,
  Hallinan, Kowalski  \& Hawley}{Berdyugina et~al.}{2017}]{Berdyugina2017}
Berdyugina S.~V.,  Harrington D.~M.,  Kuzmychov O.,  Kuhn J.~R.,  Hallinan G.,
  Kowalski A.~F.,   Hawley S.~L.,  2017, \mn@doi [Astrophys. J.]
  {10.3847/1538-4357/aa866b}, 847, 61

\bibitem[\protect\citeauthoryear{Braun}{Braun}{1970}]{Braun1970}
Braun M.,  1970, \mn@doi [J. of Differential Equations]
  {10.1016/0022-0396(70)90009-4}, 8, 294

\bibitem[\protect\citeauthoryear{Cerutti \& Beloborodov}{Cerutti \&
  Beloborodov}{2017}]{Cerutti2017}
Cerutti B.,  Beloborodov A.~M.,  2017, \mn@doi [Space Science Reviews]
  {10.1007/s11214-016-0315-7}, 207, 111

\bibitem[\protect\citeauthoryear{Chen \& Beloborodov}{Chen \&
  Beloborodov}{2014}]{Chen2014}
Chen A.~Y.,  Beloborodov A.~M.,  2014, \mn@doi [\aplett]
  {10.1088/2041-8205/795/1/L22}, 795, L22

\bibitem[\protect\citeauthoryear{Deutsch}{Deutsch}{1955}]{Deutsch1955}
Deutsch A.~J.,  1955, Ann. d'Astrophys., 18, 1

\bibitem[\protect\citeauthoryear{Dragt}{Dragt}{1965}]{Dragt1965}
Dragt A.~J.,  1965, \mn@doi [Rev. Geophys.] {10.1029/RG003i002p00255}, 3, 255

\bibitem[\protect\citeauthoryear{Dullin, Hor\'{a}nyi  \& Howard}{Dullin
  et~al.}{2002}]{Dullin2002}
Dullin H.,  Hor\'{a}nyi M.,   Howard J.,  2002, \mn@doi [Physica D]
  {http://doi.org/10.1016/S016727890200550-X}, 171, 178

\bibitem[\protect\citeauthoryear{Epp \& Masterova}{Epp \&
  Masterova}{2013}]{Epp2013effective}
Epp V.,  Masterova M.,  2013, \mn@doi [\apss]
  {http://dx.doi.org/10.1007/s10509-013-1415-4}, 345, 315

\bibitem[\protect\citeauthoryear{Epp \& Masterova}{Epp \&
  Masterova}{2014}]{EppMasterova2014}
Epp V.,  Masterova M.,  2014, \mn@doi [\apss]
  {http://dx.doi.org/10.1007/s10509-014-2066-9}, 353, 473

\bibitem[\protect\citeauthoryear{Fitzpatrick \& Mestel}{Fitzpatrick \&
  Mestel}{1988}]{Fitzpatrick}
Fitzpatrick R.,  Mestel L.,  1988, \mn@doi [\mnras] {10.1093/mnras/232.2.277},
  232, 277

\bibitem[\protect\citeauthoryear{{Goldreich} \& {Julian}}{{Goldreich} \&
  {Julian}}{1969}]{Goldreich1969ApJ}
{Goldreich} P.,  {Julian} W.~H.,  1969, \mn@doi [ApJ] {10.1086/150119}, \href
  {http://adsabs.harvard.edu/abs/1969ApJ...157..869G} {157, 869}

\bibitem[\protect\citeauthoryear{Goldstein, Poole  \& Safko}{Goldstein
  et~al.}{2002}]{Goldstein2002}
Goldstein H.,  Poole C.,   Safko J.,  2002, Classical Mechanics.
Addison Wesley

\bibitem[\protect\citeauthoryear{Graef \& Kusaka}{Graef \&
  Kusaka}{1938}]{Graef1938}
Graef C.,  Kusaka S.,  1938, \mn@doi [Journal of Mathematics and Physics]
  {10.1002/sapm193817143}, 17, 43

\bibitem[\protect\citeauthoryear{Hallinan et~al.,}{Hallinan
  et~al.}{2015}]{Hallinan2015}
Hallinan G.,  et~al., 2015, \mn@doi [Nature] {10.1038/nature14619}, 523, 568

\bibitem[\protect\citeauthoryear{I{\~{n}}arrea, Lanchares, Palaci\'{a}n,
  Pascual, Salas  \& Yanguas}{I{\~{n}}arrea et~al.}{2005}]{Inarrea2005}
I{\~{n}}arrea M.,  Lanchares V.,  Palaci\'{a}n J.~F.,  Pascual A.~I.,  Salas
  J.~P.,   Yanguas P.,  2005, \mn@doi [Phys. Let. A]
  {http://doi.org/10.1016/j.physleta.2005.02.038}, 338, 247

\bibitem[\protect\citeauthoryear{Ioanoviciu}{Ioanoviciu}{2015}]{Ioanoviciu2015}
Ioanoviciu D.,  2015, \mn@doi [SpringerPlus] {10.1186/s40064-015-0917-7}, 4,
  130

\bibitem[\protect\citeauthoryear{{Jackson}}{{Jackson}}{1976}]{Jackson1976ApJ}
{Jackson} E.~A.,  1976, \mn@doi [\apj] {10.1086/154446}, \href
  {http://adsabs.harvard.edu/abs/1976ApJ...206..831J} {206, 831}

\bibitem[\protect\citeauthoryear{Jontof-Hutter \& Hamilton}{Jontof-Hutter \&
  Hamilton}{2012}]{Jontof2012}
Jontof-Hutter D.,  Hamilton D.~P.,  2012, \mn@doi [Icarus]
  {10.1016/j.icarus.2011.09.033}, 218, 420

\bibitem[\protect\citeauthoryear{Kao, Hallinan, Pineda, Escala, Burgasser,
  Bourke  \& Stevenson}{Kao et~al.}{2016}]{Kao2016}
Kao M.~M.,  Hallinan G.,  Pineda J.~S.,  Escala I.,  Burgasser A.,  Bourke S.,
   Stevenson D.,  2016, \mn@doi [Astrophys. J.] {10.3847/0004-637X/818/1/24},
  818, 24

\bibitem[\protect\citeauthoryear{Karastergiou \& Others}{Karastergiou \&
  Others}{2015}]{Karastergiou2014}
Karastergiou A.,  Others 2015, Proc. Advancing Astrophysics with the Square
  Kilometre Array (AASKA14), AASKA14, 038

\bibitem[\protect\citeauthoryear{Landau \& Lifshitz}{Landau \&
  Lifshitz}{1976}]{Landau_I}
Landau L.~D.,  Lifshitz E.~M.,  1976, Mechanics, 3 edn.
 Course of Theoretical Physics Vol. 1, Butterworth Heinemann

\bibitem[\protect\citeauthoryear{Landau, Lifshitz  \& Pitaevskii}{Landau
  et~al.}{1984}]{Landau_8}
Landau L.~D.,  Lifshitz E.~M.,   Pitaevskii L.~P.,  1984, Electrodynamics of
  Continuous Media, 2 edn.
 Course of Theoretical Physics Vol. 8, Butterworth Heinemann

\bibitem[\protect\citeauthoryear{Michel}{Michel}{1973}]{Michel1973}
Michel F.~C.,  1973, \mn@doi [\apj] {10.1086/151956}, 180, 207

\bibitem[\protect\citeauthoryear{Michel}{Michel}{1980}]{Michel1980}
Michel F.~C.,  1980, \mn@doi [\apss] {10.1007/BF00642176}, 72, 175

\bibitem[\protect\citeauthoryear{Michel}{Michel}{1991}]{Michel1991}
Michel F.~C.,  1991, Theory of Neutron Star Magnetospheres.
University of Chicago Press

\bibitem[\protect\citeauthoryear{Ostriker \& Gunn}{Ostriker \&
  Gunn}{1969}]{Ostriker1969}
Ostriker J.~P.,  Gunn J.~E.,  1969, \mn@doi [\apj] {10.1086/150160}, 157, 1395

\bibitem[\protect\citeauthoryear{Pineda, Hallinan  \& Kao}{Pineda
  et~al.}{2017}]{Pineda2017}
Pineda J.~S.,  Hallinan G.,   Kao M.~M.,  2017, \mn@doi [Astrophys. J.]
  {10.3847/1538-4357/aa8596}, 846, 75

\bibitem[\protect\citeauthoryear{Sarychev}{Sarychev}{2010}]{Sarychev2010}
Sarychev V.~T.,  2010, \mn@doi [Radiophysics and Quantum Electronics]
  {10.1007/s11141-010-9198-8}, 52, 900

\bibitem[\protect\citeauthoryear{Schmidt}{Schmidt}{1979}]{Schmidt1979}
Schmidt G.,  1979, {Physics of high temperature plasmas}.
Academic Press

\bibitem[\protect\citeauthoryear{Schuster \& Thielheim}{Schuster \&
  Thielheim}{1987}]{Schuster1987}
Schuster R.,  Thielheim K.~O.,  1987, Journal of Physics A: Mathematical and
  General, 20, 5511

\bibitem[\protect\citeauthoryear{Shebalin}{Shebalin}{2004}]{Shebalin2004}
Shebalin J.~V.,  2004, \mn@doi [Phys. Plasmas] {10.1063/1.1752931}, 11, 3472

\bibitem[\protect\citeauthoryear{St{\o}rmer}{St{\o}rmer}{1907}]{Stormer1907}
St{\o}rmer C.,  1907, Arch. Sci. Phys. Nat., 24, 175

\bibitem[\protect\citeauthoryear{St{\o}rmer}{St{\o}rmer}{1955}]{Stormer1955}
St{\o}rmer C.,  1955, The polar aurora.
International monographs on radio, Oxford, Oxford Clarendon Press

\bibitem[\protect\citeauthoryear{Thielheim \& Wolfsteller}{Thielheim \&
  Wolfsteller}{1990}]{Thielheim1990}
Thielheim K.~O.,  Wolfsteller H.,  1990, \mn@doi [Journal of Physics A:
  Mathematical and General] {10.1088/0305-4470/23/4/029}, 23, 593

\bibitem[\protect\citeauthoryear{Wada \& Shibata}{Wada \&
  Shibata}{2011}]{Wada2011}
Wada T.,  Shibata S.,  2011, \mn@doi [\mnras]
  {10.1111/j.1365-2966.2011.19510.x}, 418, 612

\makeatother
\end{thebibliography}





\bsp	
\label{lastpage}
\end{document}